\documentclass[%
 reprint,
 superscriptaddress,
 amsmath,amssymb,
 aps,
 prb,
]{revtex4-2}

\usepackage{graphicx}% Include figure files
\usepackage{dcolumn}% Align table columns on decimal point
\usepackage{bm}% bold math
\usepackage{dsfont}% special shape latters
\usepackage{braket}% Dirac notation
\usepackage{hyperref}% add hypertext capabilities
\usepackage{xcolor}
\usepackage{soul}
\usepackage{changes}
\usepackage{ulem}

\newcommand{\RNum}[1]{\uppercase\expandafter{\romannumeral #1\relax}}% for Roman number
\let\vec\mathbf% vector as bold form

\begin{document}

\title{Localized Plasmons in Topological Insulators}

\author{Zhihao Jiang}
\email{zhihaoji@usc.edu}
\affiliation{Department of Physics and Astronomy, University of Southern California, Los Angeles, California 90089-0484, USA}

\author{Malte R\"{o}sner}
\affiliation{Institute for Molecules and Materials, Radboud University, Heyendaalseweg  135, 6525 AJ Nijmegen, The Netherlands}

\author{Roelof E. Groenewald}
\email{current address: Modern Electron LLC, Bothell, Washington 98011, USA}
\affiliation{Department of Physics and Astronomy, University of Southern California, Los Angeles, California 90089-0484, USA}

\author{Stephan Haas}
\affiliation{Department of Physics and Astronomy, University of Southern California, Los Angeles, California 90089-0484, USA}

\date{\today}

%******** abstract ********
\begin{abstract}
We investigate in a fully quantum-mechanical manner how the many-body excitation spectrum of topological insulators is affected by the presence of long-range Coulomb interactions. In the one-dimensional Su-Schrieffer-Heeger model and its mirror-symmetric variant strongly localized plasmonic excitations are observed which originate from topologically non-trivial single-particle states. These \textit{``topological plasmons"} inherit some of the  characteristics of their constituent topological single-particle states, but they  are not equally well protected against disorder due to the admixture of non-topological bulk single-particle states in the polarization function.  The strength of the effective Coulomb interactions is also shown to have strong effects on the plasmonic modes. Furthermore, we show how external modifications via dielectric screening and applied electric fields with distinct symmetries can be used to study  topological plasmons, thus allowing for experimental verification of our atomistic predictions. 
\end{abstract}

\maketitle

\begin{figure*}
 \includegraphics[width=0.95\textwidth]{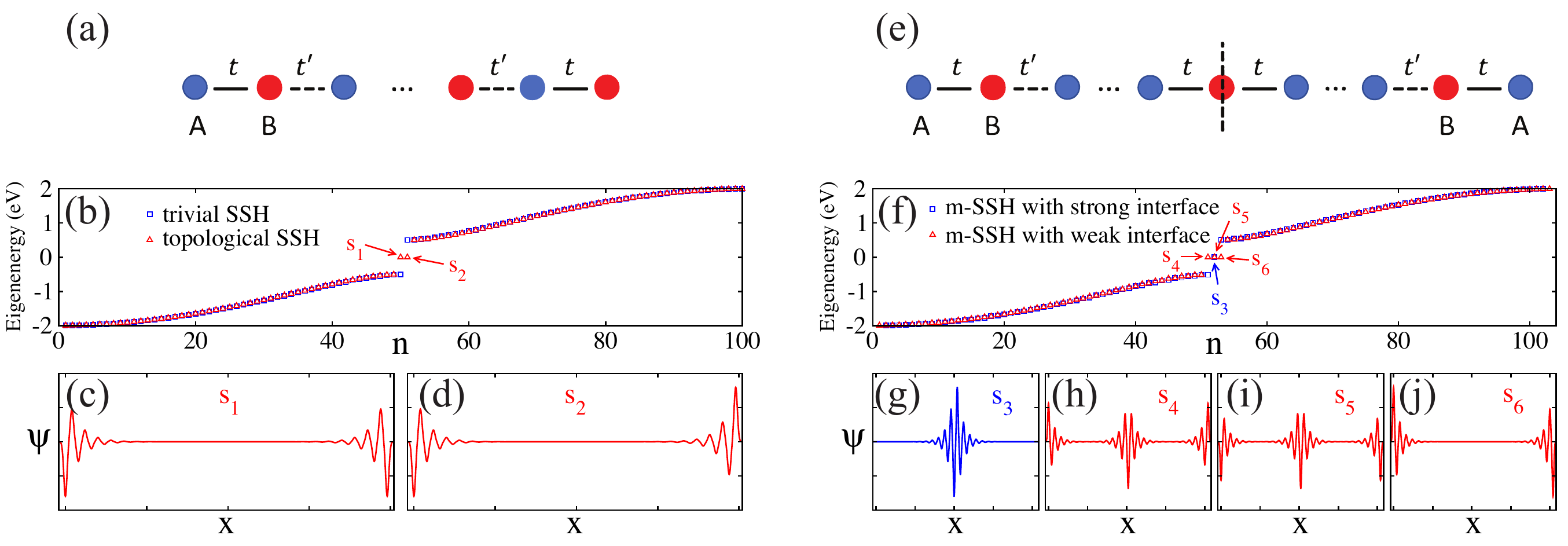}
 \caption{(a) Illustration of the simple SSH model on a bipartite tight-binding chain. (b) Corresponding energy levels of the SSH model in a 100-site open chain. We observe two zero-energy edge states ($s_1$ and $s_2$) in the topological phase, whereas no zero-energy edge states are present in the trivial phase. (c-d) Wave functions of the zero-energy states $s_1$ and $s_2$, showing localization at the edges. (e) Illustration of a mirror SSH model with inversion at the center of the chain. (f) Corresponding energy levels of the m-SSH model on a 103-site open chain. We observe localized zero-energy interface and edge states ($s_{3-6}$), depending on the interface and edge properties of the chain. (g-j) Wave functions of the localized interface state $s_3$ and localized interface and edge states $s_{4,5,6}$.}\label{Models} 
\end{figure*}

%******** section-1 Introduction ********
\section{\label{sec-1Intro} Introduction}

The experimental observation of the integer \cite{klitzing_new_1980} and fractional \cite{tsui_two-dimensional_1982} quantum Hall effects ultimately lead to the discovery of fundamentally new topologically non-trivial quantum phases \cite{hasan_colloquium:_2010,kosterlitz_nobel_2017,haldane_nobel_2017}. These manifest themselves in systems with gapped bulk states and symmetry protected conducting surface states. They can be realized, for example, in insulating materials, which are called topological insulators (TI). TIs have been found in 3D \cite{zhang_topological_2009} and 2D \cite{bernevig_quantum_2006} materials as well as in 1D meta-materials \cite{bleckmann_spectral_2017}. These systems have been characterized in great detail, with a focus on their single-particle electronic properties \cite{chen_symmetry_2013}, but the role of many-body interactions has so far been widely neglected. Just recently some attention has been given to the effects of electron-electron (Coulomb) interactions in these materials with regards to their plasmonic excitations. These many-body excitations describe collective oscillations of the electronic sea resulting from the long-range Coulomb interactions. In 3D TIs the topological surface states are formed by massless Dirac fermions which host plasmonic excitations \cite{di_pietro_observation_2013,politano_interplay_2015,politano_optoelectronic_2017,mondal_coherent_2018}, similar to those found in graphene. These have theoretically been described on a macroscopic \cite{ginley_coupled_2018} as well as on fully quantum-mechanical microscopic level using the random phase approximation (RPA) \cite{stauber_plasmonics_2017}. In two and one dimensions topological meta-materials have been created, which host 1D \cite{politano_interplay_2015,gao_probing_2016} and 0D \cite{cheng_topologically_2015,ling_topological_2015,bleckmann_spectral_2017,pocock_topological_2018,downing2018topological} plasmonic excitations. Such collective modes in these systems have so far been described mostly on a macroscopic level.

1D TIs are thereby of particular interests due to their non-continuous edge-state energie spectra. In contrast to 2D or 3D TIs, which show continuous topological conducting edge or surface bands, 1D TIs are characterized by degenerate zero-energy single-particle edge states. As we will show in the following, collective \textit{topological plasmonic excitations} in 1D TIs can  originate only from virtual excitations between topologically trivial and non-trivial states, whereas plasmonic excitations from topologically non-trivial states only are forbidden. This admixture of topologically different states renders the localized plasmonic modes in 1D TIs special, as we will show in detail in the following for the 1D Su-Schrieffer-Heeger (SSH) model \cite{su1979solitons}. 
Specifically, we calculate the real-space modulations of plasmons, from which we observe localized modes only when the 1D system is in a topologically non-trivial phase. We show in detail how these localized plasmons, which originate from constituent topological electronic states, are affected by bulk electronic states.  We call them \textit{topological plasmons} in this paper to emphasize their specific origin and interpret them as topological features in the spectrum of collective excitations. We present an in-depth study of their robustness against disorder, whereby we find relatively stable real-space excitation patterns, but strongly varying plasmonic excitation energies. Furthermore, we investigate how these topological plasmon modes are affected by Coulomb interactions and how they can be externally tuned by specifically shaped electromagnetic fields. 

The remainder of this paper is organized in the following way. In \autoref{sec-2Model}, we introduce the models under consideration, and briefly review their topological properties, before we discuss the real-space RPA method in \autoref{sec-3Method}. In \autoref{sec-4Result} we present our main results, including the observation of localized plasmon modes of topological origin, the robustness of these modes against disorder, the effects of Coulomb interactions on these modes and the excitation of these modes subject to different external fields. This is followed by conclusions in \autoref{sec-5Conclu}.

%******** section-2 models ********

\section{\label{sec-2Model} Models}

We start with the simplest one-dimensional topological insulator, the SSH model \cite{su1979solitons}, whose Hamiltonian can be written as
\begin{align}
\hat{H} =   t & \sum_{m=1}^N (\ket{m,B}\bra{m,A} + h.c.) \nonumber \\
          + t'& \sum_{m=1}^{N-1} (\ket{m+1,A}\bra{m,B} + h.c.),
\end{align}    
where $N$ is the number of unit cells, and $A$ and $B$ label the two-atomic sub-lattices. Respectively, $t$ and $t'$ describe intra- and inter-cell hopping [Fig.~\ref{Models}(a)]. For periodic boundary conditions, we can write this Hamiltonian in momentum space as \cite{asboth2016short}
\begin{align}
 \hat{H}(k) = 
 \begin{pmatrix}
  0 & h^*(k) \\
  h(k) & 0 \\
 \end{pmatrix}, \label{H_k_SSH}
\end{align}
where $h(k)=h_x(k)+ih_y(k)$, $h_x(k) = \mathrm{Re}(t) + |t'|\cos[ka+\arg(t')]$, $h_y(k) = -\mathrm{Im}(t) + |t'|\sin[ka+\arg(t')]$, and $a$ is the lattice spacing. The bulk topological invariant of the SSH model is the winding number $\mathcal{W}$, which can be obtained via \cite{asboth2016short}
\begin{align}
 \mathcal{W} = \frac{1}{2\pi i} \int_{-\pi}^{\pi}dk \frac{d}{dk} \ln [h(k)].
\end{align}
Throughout this paper we choose $t$ and $t'$ to be real numbers and set $a=2\mbox{\AA}$. For $t>t'$ we obtain $\mathcal{W}=0$, and the system is correspondingly in the trivial phase, whereas for $t<t'$ the winding number is non-zero, $\mathcal{W}=1$, and the system is in the topological phase. The phase transition occurs at $t=t'$, i.e. where the bulk band gap closes. 

Bulk-boundary correspondence implies that the topology of the SSH model can also be recognized by the number of zero-energy edge states  $N_\text{es} = 2\mathcal{W}$ in the case of open boundary conditions. Fig.~\ref{Models}(b) shows the energy spectrum of the trivial ($t=1.25\ \mathrm{eV} > 0.75\ \mathrm{eV} = t'$) and topological ($t=0.75\ \mathrm{eV} < 1.25\ \mathrm{eV} = t'$) SSH models with $100$ sites. The electronic structure in the trivial phase corresponds to a gapped particle-hole symmetric insulator with $N_\text{es}=0$, whereas in the topological phase we observe $N_\text{es}=2$, i.e. we find two degenerate zero-energy electronic states, denoted as $s_1$ and $s_2$ in Fig.~\ref{Models}(b), in the center of the band gap. Their wave functions are localized at the edges of the chain, as shown in Fig.~\ref{Models}(c) and Fig.~\ref{Models}(d). The appearance of these two zero-energy edge states is a result of the chiral symmetry of the SSH model. In this case, the two zero-energy edge states are chiral partners of each other. 

Additionally, we construct a variant of the SSH model by reflecting the simple SSH chain at one edge site. This model is mirror-symmetric, with an interface in the center [Fig.~\ref{Models}(e)], which we call the mirror-SSH (m-SSH) model in the remainder of this paper. The interface connecting two topologically distinct sub-SSH chains  supports an additional localized zero-energy state. This kind of topological zero-energy mode was first found by Jackiw and Rebbi \cite{jackiw1976solitons}, and is called the Jackiw-Rebbi mid-gap state. By construction, strong $t>t'$ (weak $t<t'$) edges correspond to a strong (weak) interface. In Fig.~\ref{Models}(f) we show the energy spectrum of the m-SSH model with $103$ sites for both scenarios. In both cases, we observe zero-energy states in the center of the band gap. For $t>t'$, there is only one zero-energy state [$s_3$ in Fig.~\ref{Models}(f)], which is localized at the interface [Fig.~\ref{Models}(g)]. For $t<t'$, there are three zero-energy states [$s_4$, $s_5$ and $s_6$ in Fig.~\ref{Models}(f)], which are localized at the interface and at the edges of chain [Fig.~\ref{Models}(h-j)].
\footnote{The number of edge and interface states can also be understood in terms of valence-bond decorations corresponding to the different states. For the simple SSH model in the topologically trivial state [blue in Fig.~\ref{Models}(b)], the resulting valence bond solid  connects every site with a neighbor, resulting in a perfect product state of valence bonds. In contrast, in the valence-bond decoration corresponding to the topological phase [red in Fig.~\ref{Models}(b)], the two outer bonds remain uncoupled, i.e. dangling bonds, leading to the observed $s_1$ and $s_2$ mid-gap zero-energy states. For the m-SSH model with neighboring two strong bonds at the central mirror interface [blue in Fig.~\ref{Models}(d)], the resulting three-site strongly coupled object is a non-bonding state, whereas the remaining states are valence bonds. This results in a localized state at the interface. Finally, in the opposite decoration of the m-SSH chain [red in Fig.~\ref{Models}(d)], there are dangling bonds at each end of the chain as well as at the mirror interface, resulting in localized zero-energy states at each of these positions.}

%******** section-3 methods ********
\section{\label{sec-3Method} Method}

In order to study plasmonic excitations of these models, we derive the electron energy loss spectrum (EELS) from the dielectric function
\begin{align}
	\bm{\varepsilon}(\omega) = \bm{\mathds{I}} - \bm{V} \bm{\chi}_0(\omega), \label{Eps_RPA}
\end{align} 
evaluated in the atomic basis. Here, $V$ is the density-density Coulomb interaction whose matrix elements are given by
\begin{align}
	V_{ab} =
	\begin{cases}
	e^2/(\varepsilon_b|\vec r_a - \vec r_b|) & \quad \text{if } a \neq b, \\
	U_0 / \varepsilon_b & \quad \text{if } a = b.
	\end{cases} \label{eqn:VR}
\end{align}
$\varepsilon_b$ is the background dielectric constant, $e$ the elementary charge, and $U_0 = \int d\vec r d\vec{r'} e^2 |\phi(\vec r)|^2 |\phi(\vec {r'})|^2 / |\vec{r-r'}|$ is the on-site Coulomb interaction evaluated from the 2D atomic basis function given by $\phi(\vec r) = (\sigma \sqrt{\pi})^{-1} e^{-r^2 / 2\sigma^2}$ (Gaussian orbitals) using a variance of $\sigma = 1\,$\AA. From this we obtain $U_0 = 17.38\,$eV. $\bm{\chi}_0(\omega)$ is the matrix representation of the non-interacting charge susceptibility, whose matrix elements in the random phase approximation \cite{nozieres1958electron,pines2018elementary} are given by \cite{westerhout2018plasmon} 
\begin{align}
	\left[\bm{\chi}_0(\omega)\right]_{ab} = 2 \sum_{i,j} \frac{f(E_i)-f(E_j)}{E_i-E_j-\omega-i\gamma} \psi_{ia}^*\psi_{ib}\psi_{jb}^*\psi_{ja}, \label{Chi^0}
\end{align} 
with $a$ and $b$ labeling atomic positions, $\gamma = 0.01\,\mathrm{eV}$ is a finite broadening, and $E_i$, $f(E_i)$, and $\psi_{ia}$ are the $i$-th electronic eigenenergy, the corresponding Fermi function, and the tight-binding wave function expansion coefficient of the atomic orbital $\phi_a$, respectively, as obtained from diagonalization of the (m-)SSH Hamiltonian. 

To extract the macroscopic $\text{EELS}(\omega) = -\mathrm{Im}\left[1/\varepsilon_n(\omega)\right]$, we follow the approach from Refs.~\onlinecite{westerhout2018plasmon} and \onlinecite{wang2015plasmonic}. Here, $\varepsilon_n(\omega)$ is defined as the eigenvalue of $\bm{\varepsilon}(\omega)$ which maximizes $\text{EELS}(\omega)$. The corresponding eigenvector yields a \textit{qualitative} spatial representation of the induced charge-density distribution, represented in the atomic basis. Using this definition, $\text{EELS}(\omega)$  shows local maxima at every possible plasmonic excitation energy. To obtain \textit{quantitative} excitation spectra and correspondingly induced charge distributions $\bm{\rho}_{ind}(\omega)$ in the atomic basis resulting from specific external excitations $\bm{\phi}_{ext}(\omega)$, we also evaluate
\begin{align}
	\bm{\rho}_{ind}(\omega) = \bm{\chi}(\omega) \bm{\phi}_{ext}(\omega) \label{rhoind}
\end{align}
utilizing the interacting charge susceptibility defined by \cite{wang2015plasmonic}
\begin{align}
	 \bm{\chi}(\omega) = \left[\bm{\mathds{I}} - \bm{\chi}_0(\omega)\bm{V}\right]^{-1}\bm{\chi}_0(\omega). \label{ChiRPA}
\end{align}
To obtain the induced charge distribution $\rho_{ind}(\vec r,\omega)$, we transform $\bm{\rho}_{ind}(\omega)$ from the atomic basis representation to the $\vec r$-space representation according to
\begin{align}
    \rho_{ind}(\vec r,\omega) = \sum_a \left[\rho_{ind}(\omega)\right]_{a} \phi_a(\vec r)
\end{align}
with $\phi_a(\vec r)$ the atomic orbital centered at site $a$. 
The induced charge generates the induced potential $\phi_{ind}(\vec r,\omega) = \int d\vec{r'} \rho_{ind}(\vec{r'},\omega)/|\vec{r-r'}|$, leading to spatially distributed induced electric field $\vec E_{ind}(\vec r, \omega) = - \nabla \phi_{ind}(\vec r,\omega)$. The corresponding excitation spectrum is finally obtained from the frequency-dependent induced energy $U_{ind}(\omega) = \int |\vec E_{ind}(\vec r, \omega)|^2 d\vec r$ \cite{muniz2009plasmonic}. As in the EELS, the plasmon frequencies maximize $U_{ind}(\omega)$.

In contrast to $\text{EELS}(\omega)$, $U_{ind}(\omega)$ and $\rho_{ind}(\vec r,\omega)$ depend on the actually applied electromagnetic field $\bm{\phi}_{ext} (\vec r, \omega)$, and result in quantitative induced charge densities, induced electric fields and induced energies, allowing for direct comparisons to experiments. Throughout this paper we analyze both, $\text{EELS}(\omega)$ and $U_{ind}(\omega)$, depending on the specific purpose of the calculation. 

%******** section-4 results ********
\section{\label{sec-4Result} Results}

\subsection{\label{sec4-1LocalPlas} Plasmonic Excitations in the SSH and m-SSH Models}

\begin{figure}[htbp]
 \includegraphics[width=8.6cm]{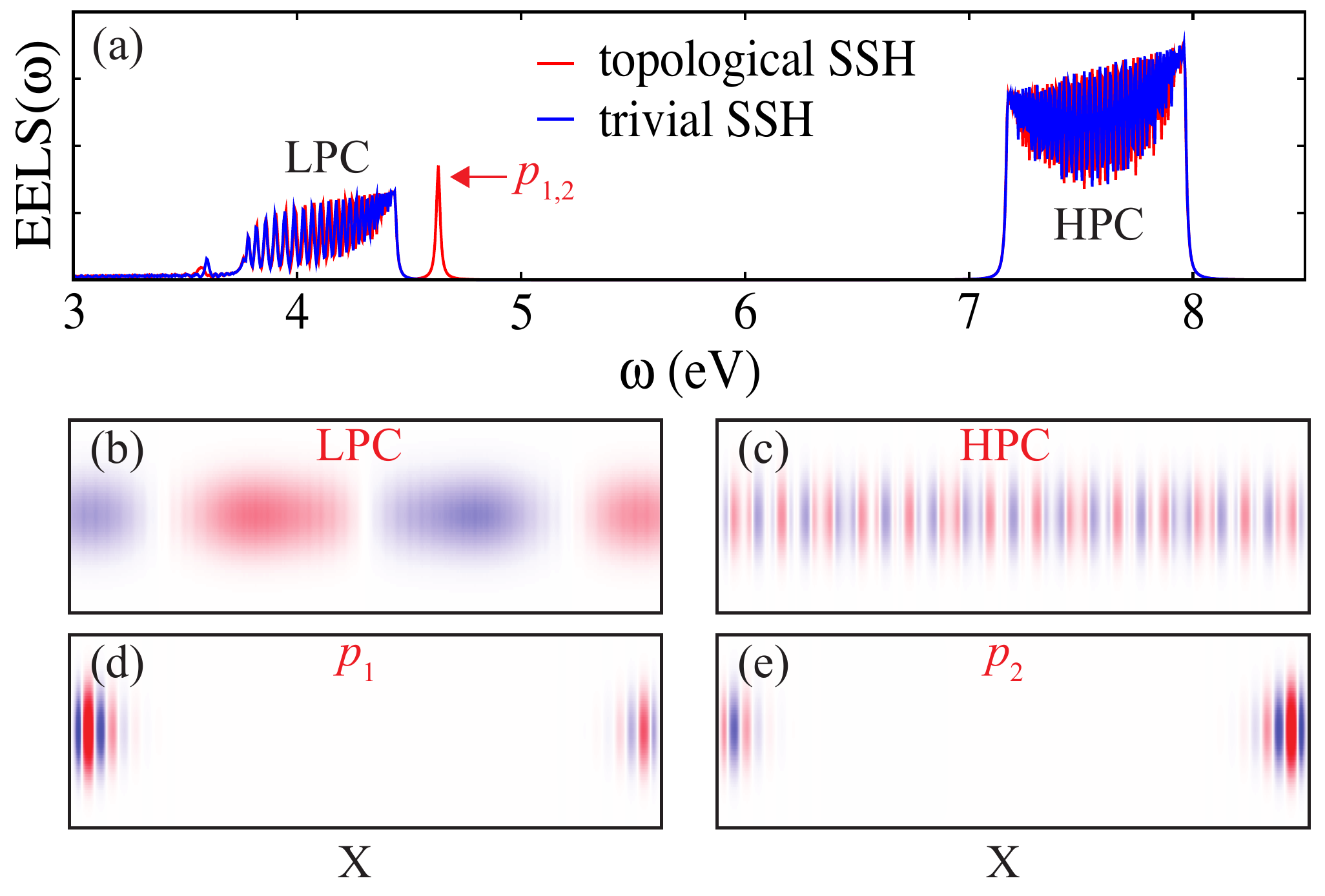}
 \caption{(a) Electron Energy Loss Spectrum (EELS) of the topologically trivial (blue) and non-trivial (red) SSH models on an open-ended 100-site chain. (b) Typical charge modulation of a bulk plasmon in the lower plasmon continuum (LPC). (c) Typical charge modulation of a bulk plasmon in the higher plasmon continuum (HPC). (d) and (e) are the charge modulations of the two-fold degenerate localized plasmon, indicated by the red arrow ($p_{1,2}$) in (a), which is only observed in the topological phase.}\label{EELS_SSH}
\end{figure}

We begin by examining the plasmonic modes in the SSH model with open boundaries. Fig.~\ref{EELS_SSH}(a) shows the corresponding $\text{EELS}(\omega)$ in the trivial ($t=1.25, t'=0.75$) and topologically non-trivial ($t=0.75, t'=1.25$) phases. In both phases, we find a low- and a high-energy plasmonic continuum (LPC/HPC) which are separated by a plasmonic gap. These continua are bulk properties, which we can study in the momentum space for a better understanding of their origination (see Appendix~\ref{Eps_RPA_q}). From this we find that these two continua result from the two internal degrees of freedom (sub-lattices $A$ and $B$) in the unit cell. Plasmons in the LPC (HPC) have inter- (intra-) unit-cell charge modulations, showing longer (shorter) oscillation wavelengths over the entire chain, as depicted in Fig.~\ref{EELS_SSH}(b) [Fig.~\ref{EELS_SSH}(c)]. While the lower bound of the LPC is entirely inherited from the single-particle band gap, the gap between LPC and HPC depends on both the single-particle band gap and the Coulomb interaction details.

In the topologically non-trivial phase, we find an additional excitation at $\omega \approx 4.63\ \mathrm{eV}$, indicated by the red arrow in Fig.~\ref{EELS_SSH}(a). It is a two-fold degenerate collective mode whose real-space charge-distribution pattern can be either even [$p_1$, Fig.~\ref{EELS_SSH}(d)] or odd [$p_2$, Fig.~\ref{EELS_SSH}(e)]. In either case, the charge distribution is highly localized at the edges of the SSH chain. 

\begin{figure}[htbp]
 \includegraphics[width=8.6cm]{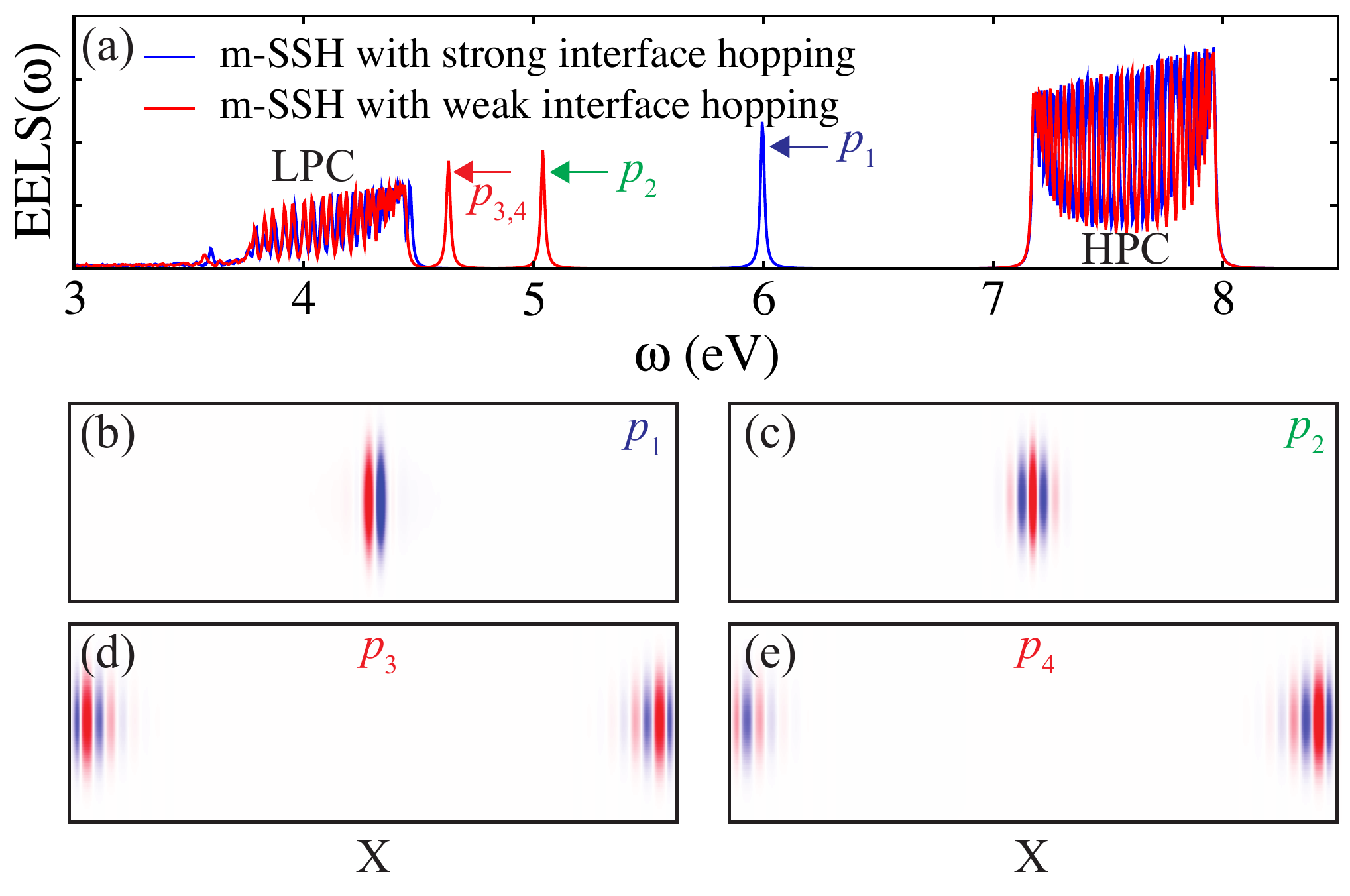}
 \caption{(a) EELS of the m-SSH model on an open-ended 103-site chain with strong (blue) and weak (red) hopping at the central mirror interface. (b) and (c) show the charge modulation of the localized interface plasmon in the m-SSH model with strong and weak interface hopping, indicated by the blue ($p_1$) and green ($p_2$) arrows in (a). (d) and (e) are the charge modulations of the two-fold degenerate localized edge plasmon, indicated by the red arrow ($p_{3,4}$) in (a), observed in the m-SSH model with weak interface hopping.}\label{EELS_mSSH}
\end{figure}

Fig.~\ref{EELS_mSSH}(a) shows the $\text{EELS}(\omega)$ of the m-SSH model with strong ($t=1.25, t'=0.75$) and weak ($t=0.75, t'=1.25$) interface hopping parameters, corresponding to topologically distinct states. In both cases, we again observe a LPC, a HPC as well as additional localized plasmonic excitations, labeled by $p_{1,2,3,4}$. In the strong interface case, there is just one additional mode $p_1$ at $\omega_1 \approx 6.0\ \mathrm{eV}$, with a dipole-like charge distribution localized at the interface, as shown in Fig.~\ref{EELS_mSSH}(b). In the weak interface case, we find two additional plasmonic excitations, $p_2$ and $p_{3,4}$, at $\omega_{2} \approx 5.04\ \mathrm{eV}$ and $\omega_{3,4} \approx 4.63\ \mathrm{eV}$, respectively. $p_2$ is an interface mode [Fig.~\ref{EELS_mSSH}(c)], but with different symmetry compared to the dipole-like mode shown in Fig.~\ref{EELS_mSSH}(b).  $p_3$ and $p_4$ are two-fold degenerate edge-localized plasmons whose real-space charge-distribution pattern can either be even [Fig.~\ref{EELS_mSSH}(d)] or odd [Fig.~\ref{EELS_mSSH}(e)]. They are essentially the same as the localized edge plasmons observed in the simple SSH model because both, the open-ended topological SSH chain and the open-ended m-SSH chain with weak interface hoppings, have weakly linked edges ($t=0.75$), leading to dangling bonds at the two ends of the open chain. 

At this point it is important to note, that all of these plasmonic modes in the (m-)SSH model have already been theoretically described \cite{ling_topological_2015,cheng_topologically_2015,bleckmann_spectral_2017,pocock_topological_2018,downing2018topological} and partially also experimentally verified \cite{cheng_topologically_2015,bleckmann_spectral_2017} in \textit{macroscopic} 1D chains. Here we, however, focus on microscopic properties on smaller length-scale forcing us to use a fully quantum-mechanical treatment \cite{goncalves_plasmon-emitter_2019}. This allows us to disentangle non-interacting single-particle states (including their partial topological character) and the resulting many-body excitations originating from the long-range Coulomb interaction.

\subsection{\label{sec4-1bLocalPlas} Topological Origin of Localized Plasmon Excitations}

Noticeable, the localized plasmonic states marked by arrows in Figs.~\ref{EELS_SSH} and~\ref{EELS_mSSH} resemble the single-particle topological states shown in Fig.~\ref{Models}. To get a deeper understanding of the origin of these localized plasmonic excitations, we decompose the full charge susceptibility (which is just the complete charge susceptibility given by Eq.~\ref{Chi^0}) $\bm{\chi}_0^\text{full} = \bm{\chi}_0^\text{topo} + \bm{\chi}_0^\text{bulk}$ into its topological and bulk contributions by separating the summation in Eq. \eqref{Chi^0} as follows:
\begin{align}
	\underbrace{\sum_{i,j} \dots}_{\Rightarrow \, \bm{\chi}_0^\text{full}} = \underbrace{
	    \sum_{i\in \mathrm{TS}}\sum_{j\notin \mathrm{TS}} \dots + 
	    \sum_{i\notin \mathrm{TS}}\sum_{j\in \mathrm{TS}} \dots}_{\Rightarrow \, \bm{\chi}_0^\text{topo}}
	    + 
	    \underbrace{
	    \sum_{i,j\notin \mathrm{TS}} \dots}_{\Rightarrow \, \bm{\chi}_0^\text{bulk}}, \label{DecomposeChi}
\end{align}
where $\mathrm{TS}$ is the set of topological zero-energy states. Due to their degeneracy, there are no virtual excitations between the topological electronic states, so that we can ignore the term $\sum_{i,j\in \mathrm{TS}}$ in the above decomposition. We call the remaining first two terms on the right hand side of Eq.~\eqref{DecomposeChi} the topological ($\bm{\chi}_0^\text{topo}$) and the third term bulk charge susceptibility ($\bm{\chi}_0^\text{bulk}$). 

\begin{figure}[htbp]
	\includegraphics[width=8.6cm]{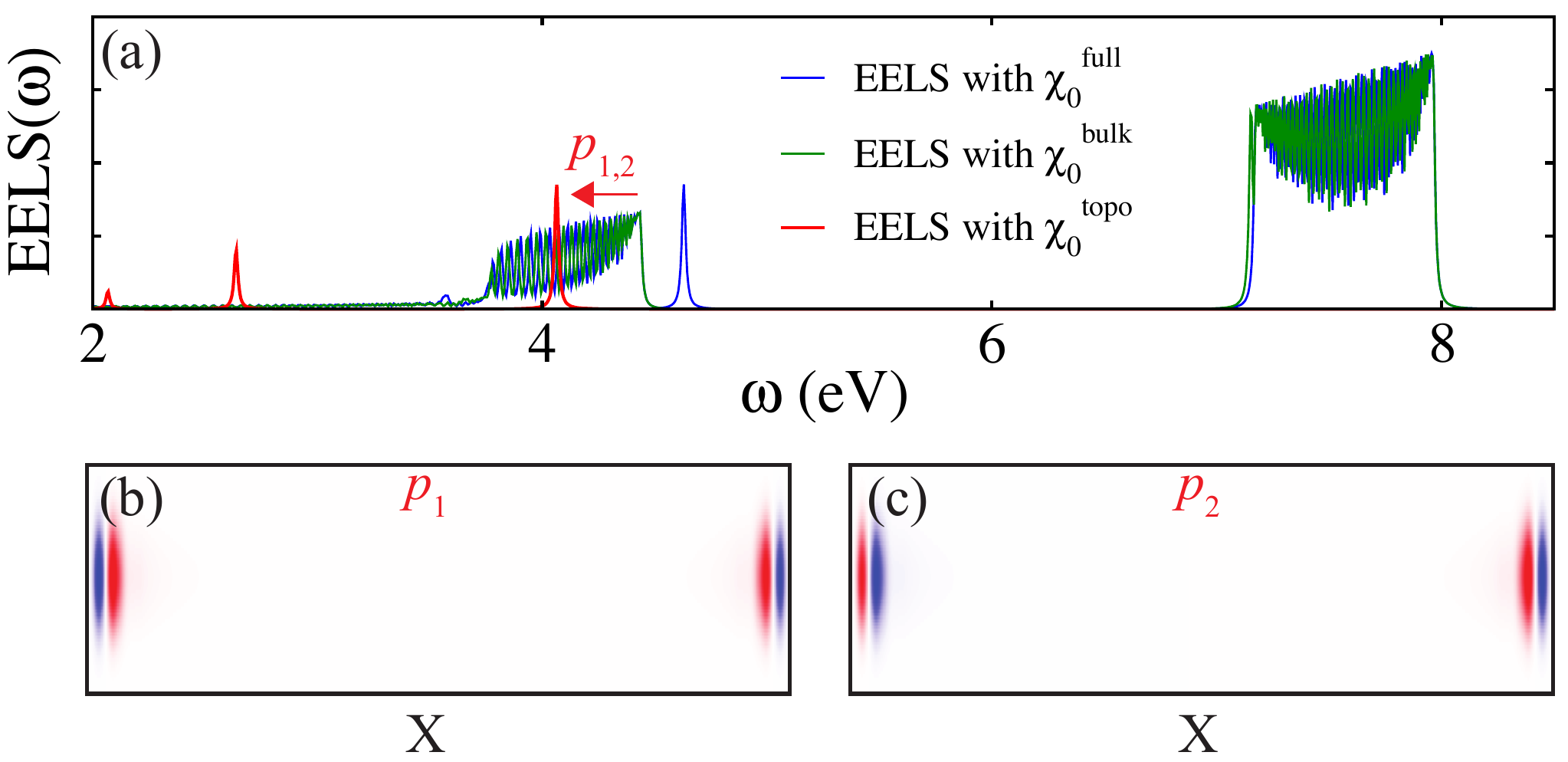}
	\caption{EELS decomposition in the SSH model. (a) EELS of the 100-site open-ended topological SSH chain, calculated using $\bm{\chi}_0^\text{full}$, $\bm{\chi}_0^\text{bulk}$ and $\bm{\chi}_0^\text{topo}$ respectively. (b) and (c) are the charge modulations of the two-fold degenerate mode at $\omega \approx 4.07\ \mathrm{eV}$ (indicated by the red arrow) observed in the topological EELS. }\label{topo_EELS}
\end{figure}

In order to delineate the bulk and edge state contributions to the plasmon spectrum, we show $\text{EELS}(\omega)$ of the topological SSH model using $\bm{\chi}_0^\text{full}$, $\bm{\chi}_0^\text{bulk}$ and $\bm{\chi}_0^\text{topo}$ in Fig.~\ref{topo_EELS}(a). As expected, $\text{EELS}^\text{bulk}(\omega)$ reproduces just the LPC and HPC, with no sign of the localized plasmon at $\omega \approx 4.63\ \mathrm{eV}$, as observed in the full $\text{EELS}(\omega)$. In contrast, $\text{EELS}^\text{topo}(\omega)$ indicates a plasmon at $\omega \approx 4.07\ \mathrm{eV}$ (indicated by the arrow) with strongly localized charge-distribution patterns [see Figs.~\ref{topo_EELS}(b) and (c)], which resemble the results shown in Figs.~\ref{EELS_SSH}(d) and (e), where the full charge susceptibility was considered. The localized plasmonic excitations in the full EELS thus originate from $\bm{\chi}_0^\text{topo}(\omega)$.

This observation becomes plausible by analyzing the matrix elements of $\bm{\chi}_0^\text{topo}$, which are given by \footnote{
In deriving Eq. \eqref{topoChi}, we assume that the tight-binding wave functions are real and all topological states have zero energy and occupation number $1/2$.
}
\begin{align}
	\left[\bm{\chi}_0^\text{topo}(\omega)\right]_{ab} = P_{ab} S_{ab}(\omega) \label{topo-Chi}
\end{align} 
with 
\begin{align}
	P_{ab} = \braket{\phi_a|\bm{P}|\phi_b} 
	       = \sum_{i \in \text{TS}}
	             \braket{\phi_a|\psi_i}
	             \braket{\psi_i|\phi_b}
\end{align} 
and
\begin{align}
	S_{ab}(\omega) = \sum_{i\notin \mathrm{TS}} \frac{E_i\left[2f(E_i) - 1\right]}{E_i^2 - (\omega+i\gamma)^2} \psi_{ia} \psi_{ib}. \label{topoChi}
\end{align}    
$\bm{P}$ is the projection operator to the space of topological zero-energy states.  Each element of $\bm{\chi}_0^\text{topo}$ is thus a product of the projection operator to the space of topological zero-energy states $P_{ab}$ and the RPA sum over all bulk electronic states $S_{ab}(\omega)$. The topological electronic states $\psi_{i \in \text{TS}}(r)$ are strongly localized at the edges, so that we find non-zero elements of $P_{ab}$ only for $a$ and $b$ close to the edges. $S_{ab}(\omega)$ does not vary abruptly with $a$ and $b$ since the bulk states extend across the entire chain. $S_{ab}(\omega)$ thus does not significantly affect the localization of the topological charge susceptibility, yielding a very sparse matrix $\bm{\chi}_0^\text{topo}$. This also holds for the matrix representation of the corresponding dielectric function $\bm{\varepsilon}^\text{topo}(\omega) = \bm{\mathds{I}} - \bm{V_c} \bm{\chi}_0^\text{topo}$, which inherits its strongly localized character from $\bm{\chi}_0^\text{topo}(\omega)$ ($\bm{V_c}$ further increases the localization due to its diagonal-dominant matrix structure).

Thus, these localized plasmonic excitations indeed originate from the localized topological electronic states, which is why we refer to them as \textit{topological plasmons} hereafter. It is important to notice that there is still a bulk-related component in $\bm{\chi}_0^\text{topo}$, as defined by Eq.~\eqref{topoChi}, as well as in the full $\bm{\chi}_0$ resulting from $\bm{\chi}_0^\text{bulk}$. These bulk contributions strongly affect the topological plasmons, i.e., they shift their excitation energy and spread out their spatial extension. We therefore expect that the localized topological plasmons are less stable than their constituent topological single-particle states.  

\begin{figure*}
 \includegraphics[width=1.95\columnwidth]{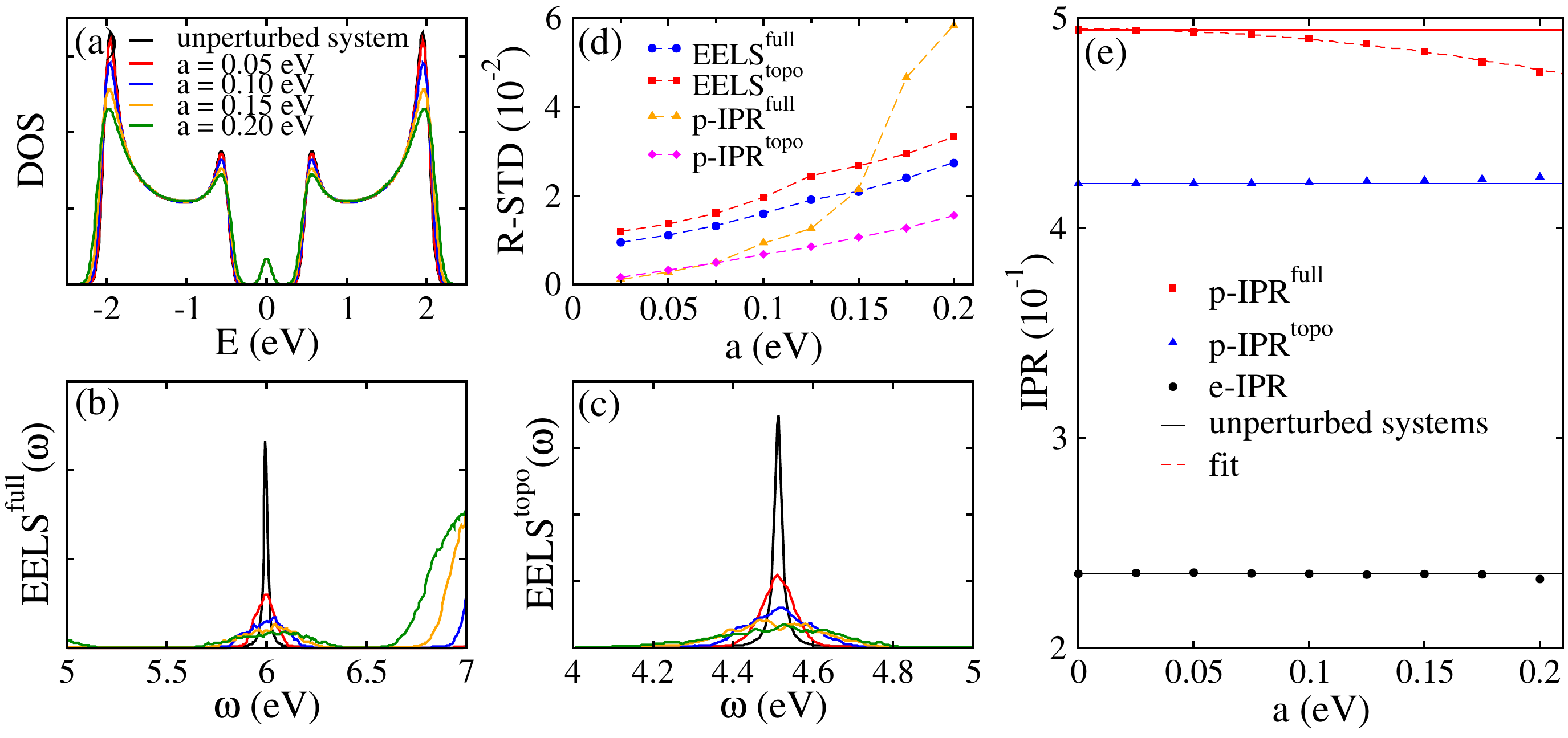}
 \caption{Open m-SSH chain with strong interface hopping, subject to uniform hopping disorder. (a) Averaged single-particle density of states over 500 realizations with different strengths of hopping disorder $a$. (b) and (c) are averaged EELS around the interface topological plasmon over 500 realizations calculated from $\bm{\chi}_0^\text{full}$ and $\bm{\chi}_0^\text{topo}$, respectively. (d) Relative standard deviation (R-STD) of the excitation energies and real-space charge modulations of the topological plasmon modes shown in (b) and (c) as functions of $a$. (e) Averaged inverse participation ratio of the topological electronic state (e-IPR) and of the interface plasmon mode calculated from $\bm{\chi}_0^\text{full/topo}$ as functions of $a$. \label{Fig_perturbation}}
\end{figure*}

\subsection{\label{sec4-2Stability} Robustness of Localized Topological Plasmons against Disorder}

We numerically study the stability of the topological plasmon localized at the interface in the m-SSH model with strong interface hoppings, i.e. the mode $p_1$ in Fig.~\ref{EELS_mSSH}(a), in the presence of uniformly distributed off-diagonal disorder $\delta t\sim \mathrm{U}(-a,a)$ in the hopping matrix elements $t$ and $t'$. The strength of the perturbation is limited to a certain range, i.e., $a<|t-t'|/2$, such that it will not induce any topological transition by reversing the order of $t$ and $t'$. We consider 500 realizations for each $a$, ranging from $0.025\ \mathrm{eV}$ to $0.2\ \mathrm{eV}$. 

These perturbations can affect the single-particle energies and wave functions as well as the plasmonic excitations and corresponding charge-modulation patterns. To study the stability of the (excitation) energies, we will compare the plasmonic electron-energy-loss-spectra with the electronic density of states. To analyze the disorder-induced changes to the plasmonic charge-modulation in comparison to the changes to the electronic wave functions, we will focus on the inverse-participation ratios (IPR\footnote{The IPR of a normalized $N$-dimensional vector $\bm{\psi}$ is calculated via $\mathrm{IPR}=\sum_i^N |\bm{\psi}_i|^4$, where $i$ is the index of the components. An increased IPR indicates a stronger localized distribution. We calculate the electronic IPR (e-IPR) from the eigenvector of the Hamiltonian and the plasmonic IPR (p-IPR) from the eigenvector of the dielectric matrix derived from $\bm{\chi}_0^\text{full/topo}$ (p-IPR$^\text{full/topo}$).}) as measures for the corresponding localization lengths.

% DOS vs. EELS
Fig.~\ref{Fig_perturbation}(a) shows the unperturbed and (averaged) perturbed electronic DOS. The bulk electronic states are affected by the perturbation, whereas the topological electronic state at zero energy is unchanged.  Fig.~\ref{Fig_perturbation}(b) shows the averaged EELS around the plasmonic interface mode, which broadens when disorder is introduced. The excitation energy of this topologically-originated plasmon is thus less stable against external perturbations than the topological single-particle state. To quantify this behaviour, we plot the relative standard deviation of the plasmonic excitation energy $\omega_p$ (R-STD\footnote{The R-STD is defined by the square-root of the second central moment of the averaged EELS, divided by the unperturbed excitation energy.}) as a function of $a$ in Fig.~\ref{Fig_perturbation}(d) (blue dots). From this we clearly see that $\omega_p$ gets more and more unstable with increasing disorder. As discussed in the previous section, we can separate the full polarization $\bm{\chi}_0^\text{full} = \bm{\chi}_0^\text{topo} + \bm{\chi}_0^\text{bulk}$ into a ``topological'' and a ``bulk'' part. From this we can calculate the ``topological'' EELS [Fig.~\ref{Fig_perturbation}(c)] and the corresponding R-STD [Fig.~\ref{Fig_perturbation}(d), red squares]. The latter shows a very similar trend of decreasing stability as discussed before. So, we can conclude that such instability of excitation energy of the topologically-originated plasmon is due to the mixed bulk electronic states inside the $\bm{\chi}_0^\text{topo}$ (Eq.~\ref{topo-Chi}). In other words, the separated bulk part of the polarization $\bm{\chi}_0^\text{bulk}$ does not drastically affect stability of the excitation energy, although it shifts the position of the excitation.

% IPRs
Next we examine the stability of the localization of the topological single-particle wave function (e-IPR) and the charge-distribution of the interface plasmon (p-IPR) as a function of $a$. The results are shown in Fig.~\ref{Fig_perturbation}(e). As expected, the e-IPR does not vary at all indicating the stability of the topological single-particle wave function. The full p-IPR (red squares) is, however, affected by the perturbation $a$. The decreasing trend in p-IPR corresponds to a decreasing localization (i.e. expansion) of the previously strongly localized interface plasmon. We fit this decay by an exponential function p-IPR$^\text{full} = A e^{-a^2/\delta}$ with $A\approx0.5$ and $\delta\approx1\,$(eV)$^2$. As $a^2$ is proportional to the variance of the uniform distribution $\mathrm{U}(-a,a)$, we find that p-IPR$^\text{full}$ decays exponentially with the variance of the perturbation.

In contrast to the plasmonic excitation energy, the stability of the plasmonic localization is strongly governed by the bulk polarization $\bm{\chi}_0^\text{bulk}$, as we can see from the comparison of full and the ``topological'' p-IPR in Fig.~\ref{Fig_perturbation}(e). The latter (blue up-triangles) neglects the bulk polarization and thereby becomes quite stable against the perturbation $a$. Thus, the disorder-induced delocalization tendency of the interface plasmon is driven by the perturbed bulk screening properties. For more details on the origin of this observation we refer to Appendix~\ref{Append_Stability}.

In order to quantify whether the plasmonic excitation energy or its charge-localization is more affected by the disorder we also plot the R-STD of the p-IPR in Fig.~\ref{Fig_perturbation}(d). From this we find that the topological plasmon localization is less affected than its excitation energy for perturbation strengths of $a \lesssim 0.15\,$eV only. 

To conclude, it is important to note that $\delta\approx1\,$(eV)$^2$ is large compared to the energy scale of the perturbation. In fact, $\sqrt{\delta}$ is here of the order of the band gap ($1 \mathrm{eV}$), which is likely beyond the limit of experimentally achievable perturbation levels. Therefore, we expect that the localization of the topologically-originated plasmon interface mode in the m-SSH model is rather stable when subjected to sufficiently small hopping noise.

\subsection{\label{sec4-3Coulomb} Effects of Coulomb Interactions on Topological Plasmons}

\begin{figure}[htbp]
 \includegraphics[width=8.6cm]{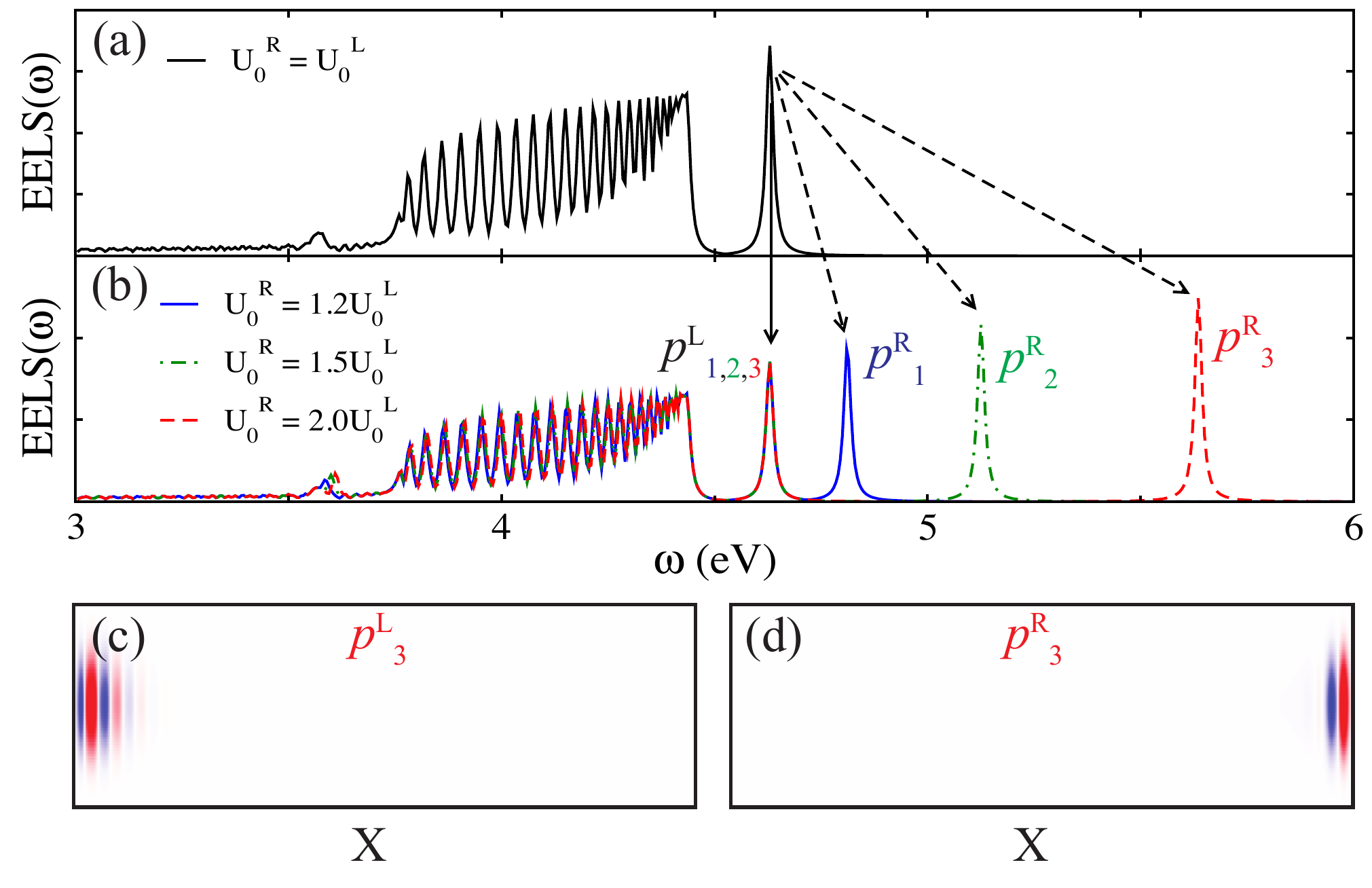}
 \caption{Removal of topological plasmon degeneracy and enhancement of plasmon localization with locally varying Coulomb interaction. (a) EELS of the SSH model with symmetric on-site Coulomb interaction $U_0^R = U_0^L$, showing a two-fold degenerate plasmon mode. (b) EELS of SSH models with asymmetric on-site Coulomb interaction $U_0^R = 1.2U_0^L, 1.5U_0^L, 2.0U_0^L$. In each case, the degenerate mode in (a) splits into a left edge mode ($p^L_{1,2,3}$, indicated by the solid arrow) and a right edge mode ($p^R_{1,2,3}$, indicated by dashed arrows). (c) and (d) are real space charge modulations of modes $p^L_3$ and $p^R_3$, showing that the right edge mode with larger on-site Coulomb interaction is more localized.} \label{degeneracy}
\end{figure}

In the topological phase of the SSH model, the topological plasmon at $\omega \approx 4.63\ \mathrm{eV}$ is a two-fold degenerate collective mode whose real-space patterns are localized at the opposite edges of the open chain [Figs.~\ref{EELS_SSH}(d) and (e)]. The single-particle topological zero-energy edge states are also two-fold degenerate. In this case, the degeneracy of single-particle topological states is fully inherited by the topological plasmons due to equivalent local Coulomb environments at the right and left edges. In the m-SSH model with weak interface hopping, we obtain one interface plasmon mode at $\omega \approx 5.04\ \mathrm{eV}$ in addition to two two-fold degenerate edge plasmon modes at $\omega \approx 4.63\ \mathrm{eV}$. The constituent single-particle topological zero-energy states, on the other hand, are three-fold degenerate, with wave functions localized at the mirror interface and at the two chain ends. In this case, the degeneracy of the constituent single-particle states is only partly inherited by the derived collective states and split into ``1+2'' in the topological plasmons. The interface mode can now be distinguished from the edge mode in the EELS due to its different local Coulomb environment, i.e., the interface site is exposed to the Coulomb interaction from the left and right part of the system, while the edges just feel one tail of the Coulomb interaction.

To verify the above argument we break the Coulomb environment equivalence of the SSH model by using different local Coulomb potentials on the left and right edges, i.e. $U_0^R \neq U_0^L$. Figs.~\ref{degeneracy}(a) and (b) show the resulting EELS$(\omega)$ with $U_0^R = U_0^L, 1.2\,U_0^L, 1.5\,U_0^L, 2\,U_0^L$. For $U_0^L \neq U_0^R$ [Fig.~\ref{degeneracy}(b)] the two-fold degenerate edge plasmon is split into two non-degenerate modes: the left mode ($p^L_{1,2,3}$) and the right mode ($p^R_{1,2,3}$) with different excitation energies. By increasing the difference between $U_0^L$ and $U_0^R$ we also increase the energy difference between the two excitation energies. With increasing on-site Coulomb interaction $U_0^R$, the charge distribution at the right edge gets more localized in contrast to the mode localized at the left edge, which, for instance, can be seen by comparing the mode $p^L_3$ [Fig.~\ref{degeneracy}(c)] and the mode $p^R_3$ [Fig.~\ref{degeneracy}(d)]. Such a manipulation of the topological plasmons by changes to the local Coulomb interactions is remarkable, since it does not affect other plasmonic excitations in the system. In contrast, global changes to the Coulomb interaction will affect all plasmons simultaneously (see Appendix~\ref{Append_GlobalCoulomb}).     

\subsection{\label{sec4-4Excitation} Excitation of Topological Plasmons Subject to Different External Potentials}

While the EELS shows all possible plasmonic excitations of a system, it does not yield any information about the excitations generated by specific external electromagnetic fields. In reality the symmetry of the external electromagnetic field will strongly influence which modes will and can be excited. Specifically, it depends on whether $\bm{\chi}_0(\omega) \bm{\phi}_{ext}(\omega)$ is zero or not [see Eqs. \eqref{rhoind} and \eqref{ChiRPA}].  We therefore turn to the induced energy spectrum $U_{ind}(\omega)$, which renders the realistic response to specific applied external electromagnetic fields.

\begin{figure}[htbp]
 \includegraphics[width=8.6cm]{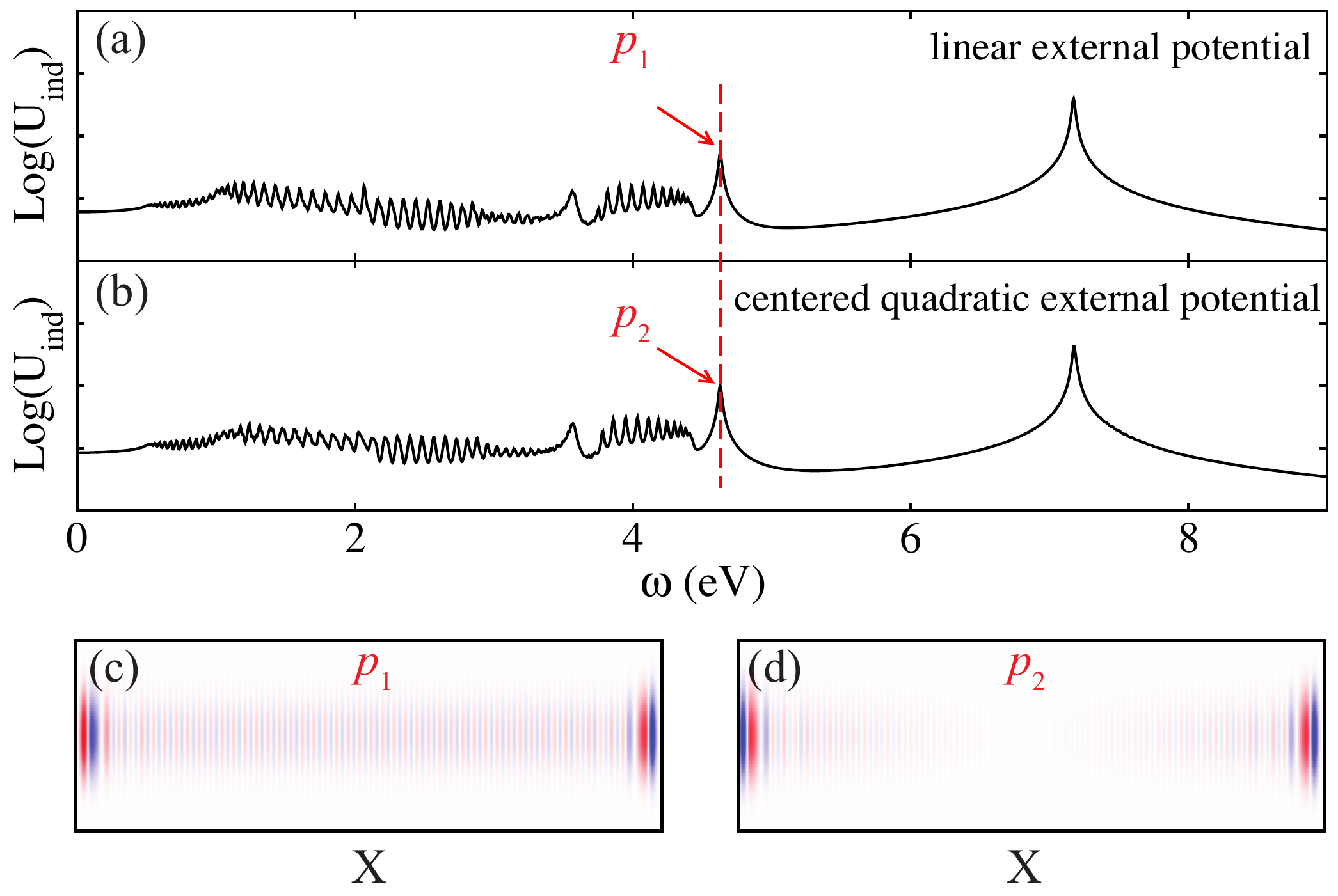}
 \caption{Induced energy in the 100-site open-ended topological SSH chain, subject to a (a) linear and to a (b) centered quadratic external electrical potential. The two-fold degenerate topological edge plasmon excitation is indicated by a red dashed line. (c) and (d) show charge-distribution patterns of this mode corresponding to the linear and the centered quadratic external potentials.} \label{excitationSSH}
\end{figure}

\begin{figure}[htbp]
 \includegraphics[width=8.6cm]{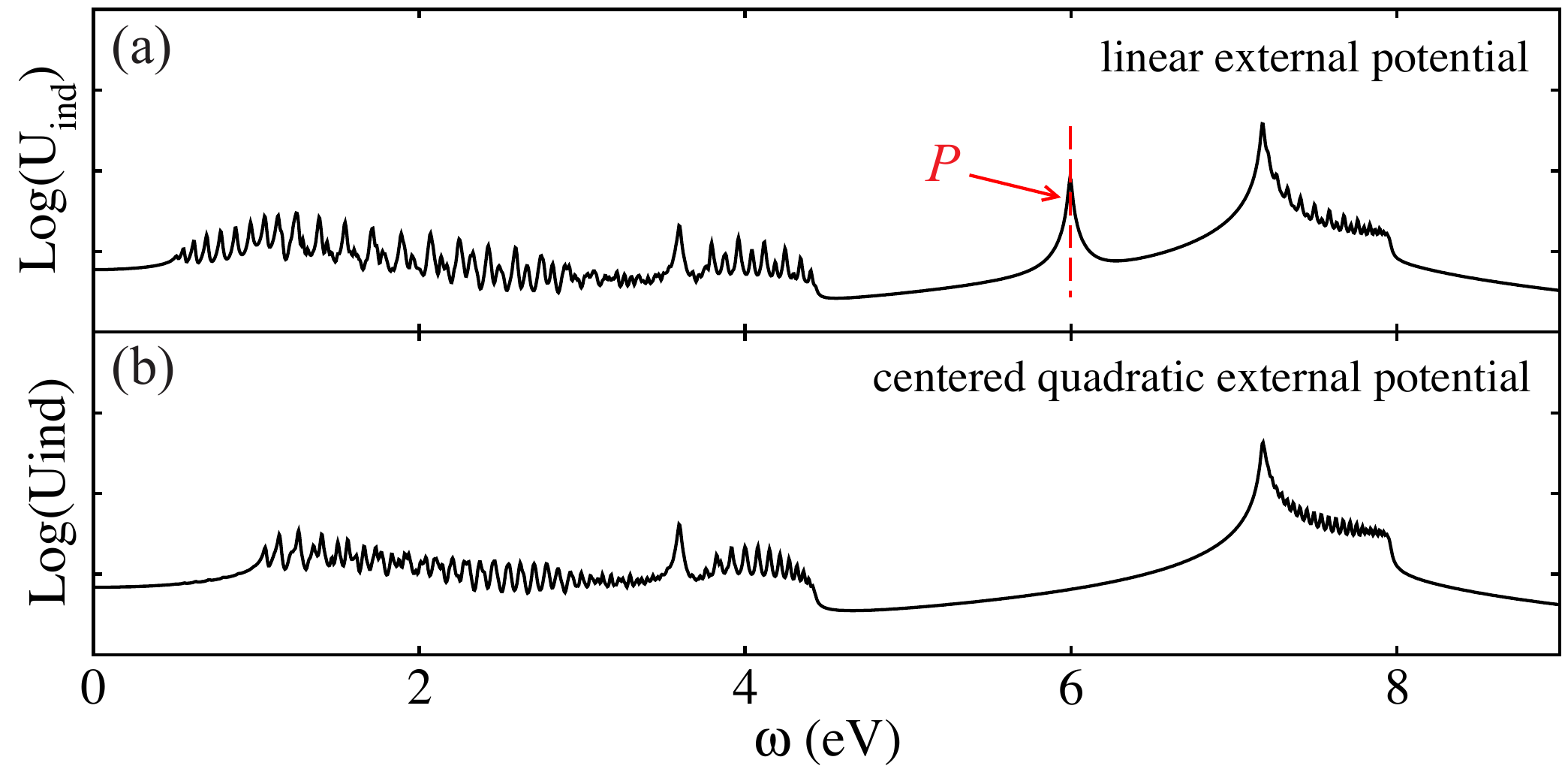}
 \caption{Induced energy in the 103-site open-ended m-SSH chain with strong interface hoppings, subject to a (a) linear and to a (b) centered quadratic external electrical potential. The topological interface plasmon excitation is indicated by a red dashed line.} \label{excitationmSSH}
\end{figure}

To this end we focus on the topological plasmons in the SSH and m-SSH models subject to linear and centered quadratic external electrical potentials. The linear potential (LP) is an odd function, whereas the centered quadratic potential (CQP) is an even function in real space. Figs.~\ref{excitationSSH}(a) and (b) show the induced energy spectra of the SSH model subject to both potentials. The topological plasmon mode at $\omega \approx 4.63\ \mathrm{eV}$, as shown in Fig.~\ref{EELS_SSH}(a), also appears in the induced energy spectrum for each of these cases (the red dashed line). Fig.~\ref{excitationSSH}(c) and (d) depict the corresponding real-space charge distributions which inherit the symmetry of the external potential. In more detail, we find an odd real-space charge distribution by using the LP [Fig.~\ref{excitationSSH}(c)] which resembles the one shown in Fig.~\ref{EELS_SSH}(e) and an even distribution by using the CQP [Fig.~\ref{excitationSSH}(d)] as found in Fig.~\ref{EELS_SSH}(d). Therefore, we can excite either one of these two degenerate modes by choosing certain external potentials.

Figs.~\ref{excitationmSSH}(a) and (b) show the induced energy spectra of the m-SSH model with strong interface hopping, also subject to LP and CQP. The EELS of this model [Fig.~\ref{EELS_mSSH}(a)] has a plasmon mode at $\omega \approx 6.00\ \mathrm{eV}$ with a dipole-like real-space charge-distribution pattern observed in Fig.~\ref{EELS_mSSH}(b). Thus, it can be excited by LP but not by CQP as verified in Fig.~\ref{excitationmSSH}.

The symmetry of the external perturbation has thus a strong effect to the topological plasmonic excitations while the rest of the spectrum is unchanged and is therefore well-suited to study these special and highly localized states.       

%******** section-5 conclusions ********
\section{\label{sec-5Conclu} Conclusions}
In summary, we comprehensively studied the plasmonic excitations in the one-dimensional SSH model and its mirror-symmetric variant (m-SSH model) using a fully quantum mechanical approach, with the Coulomb interaction considered on the RPA level. Two gapped bulk plasmonic branches as well as localized plasmonic edge states  are observed in the topologically non-trivial phases, showing resemblance to the constituent single-particle bulk energy bands and topological states. However, there are notable differences between these collective excitations  compared to the single-particle states, which have not been fully been appreciated in previous works. On one hand, due to the contribution of the bulk single-particle states in the polarization function, the stability of the localized plasmons against disorder is weakened. Furthermore, the plasmonic band gap, the excitation energies, the degeneracies, and the localization of the topologically originated plasmons are severely affected by the Coulomb interactions in the system. Remarkably, these localized plasmons can be manipulated by changing the effective local Coulomb interactions, which provides a promising tuning knob to experimentally control these modes via substrate modulation. Finally, we have provided simple guidelines to selectively excite  localized plasmons by using specifically shaped external electric fields. These findings can be used to distinguish topologically originated plasmons from bulk plasmons and to design highly stable localized plasmons in topologically non-trivial  systems. Furthermore, our RPA analysis of the Coulomb interaction can be used as the foundation for full $GW$-like calculations in topologically non-trivial systems.

\begin{acknowledgments}
This work was supported by the US Departmentof Energy grant number DE-FG03-01ER45908. M.R. and S.H. thank the Alexander-von-Humboldt Foundation for support. The numerical computations were carried out on the University of Southern California High Performance Supercomputer Cluster. We would also like to thank Yi-Zhuang You, Zhengzhi Ma, Lukas Muechler, Cyrus Dreyer, and Lorenzo Campos Venuti for useful discussions. 
\end{acknowledgments}

\appendix
\section{\label{k_SSH} Plasmon Dispersion of the SSH Model}

The RPA dielectric function of the SSH model in momentum space is given by
\begin{align}
\bm{\varepsilon}(q,\omega) = \bm{\mathds{I}} - \bm{V}(q) \bm{\chi}_0(q,\omega), \label{Eps_RPA_q}     
\end{align}
where $\bm{\varepsilon}(q,\omega)$, $\bm{V}(q)$ and $\bm{\chi}_0(q,\omega)$ are matrices in the sub-lattice basis. We calculate $\bm{V}(q)$ by Fourier transforming the real space Coulomb interaction $\bm{V}(R)$ from Eq.(\ref{eqn:VR}) evaluated on the discrete SSH lattice, as
\begin{align}
    \bm{V}_{ab}(q) = \sum_R^{}\bm{V}_{ab}(R) e^{-iqR} \label{Vq}.
\end{align}
Here, $a$ and $b$ label the atoms within a unit cell. The susceptibility matrix in the sub-lattice basis reads
\begin{align}
 [\bm{\chi}_0(q,\omega)]_{ab} = & \sum_{k,n,n'}^{}\frac{f_n(k+q)-f_{n'}(k)}{E_n(k+q)-E_{n'}(k)-\omega-i\gamma} \nonumber \\
 \
 &\psi_{n',a}^*(k) \psi_{n,a}(k+q) \psi_{n,b}^*(k+q) \psi_{n',b}(k)  \label{Chi0_q}
\end{align}
where $n$ and $n'$ are energy band indices. $E_n(k)$ and $\psi_n(k)$ of the SSH model are obtained by diagonalizing the Hamiltonian from the Eq.~\eqref{H_k_SSH}. $\psi_{n,a}(k)$ is the component of the state $\psi_n(k)$ on the sublattice $a$. The plasmon dispersions can finally be observed from $\text{EELS}(q,\omega)$, which is evaluated from the dielectric matrix $\bm{\varepsilon}(q,\omega)$ using the same method as introduced in the Sec.~\ref{sec-3Method}. 

\begin{figure}
 \includegraphics[width=1.0\columnwidth]{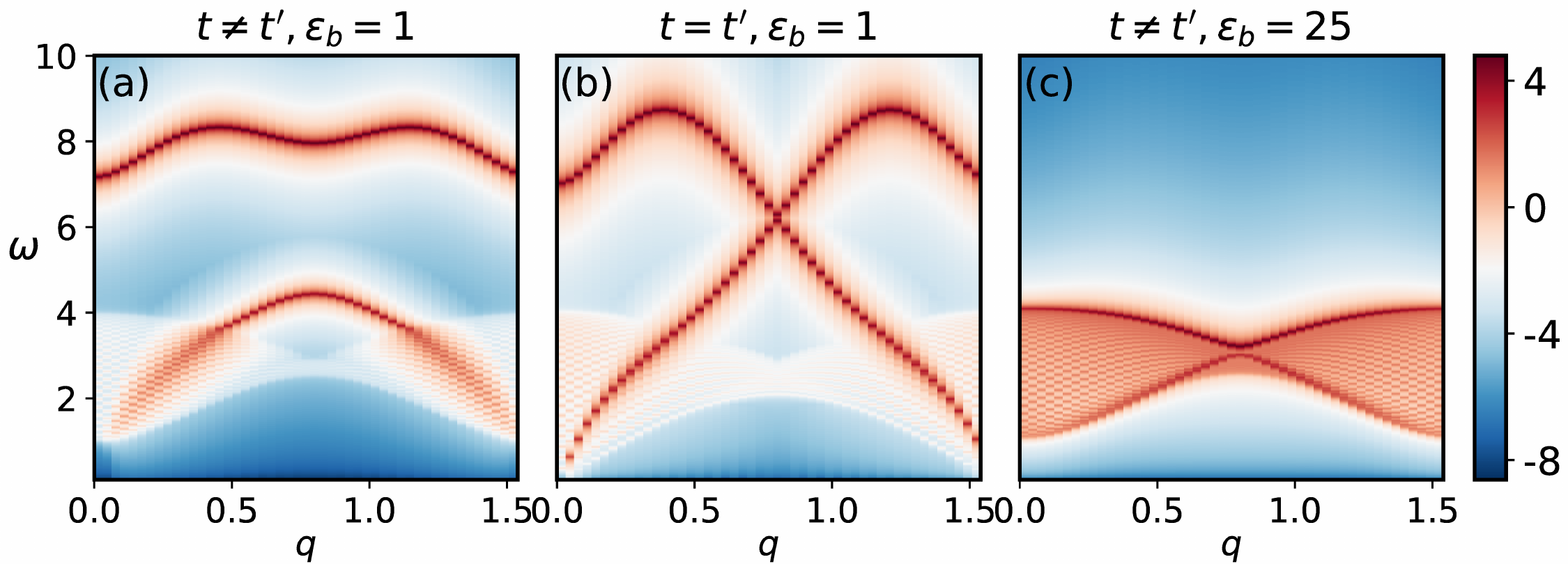}
 \caption{The EELS of the SSH model in the momentum space evaluated using different hopping parameters $t$ and $t'$ and the environmental dielectric constant $\varepsilon_b$. Specifically, shown in (a) are two plasmon branches separated by a gap, when $t\neq t'$ and $\varepsilon_b=1$. In (b), the plasmonic gap closes due to the closure of the single-particle energy gap ($t=t'$). In (c), the plasmonic gap greatly shrinks with increased environmental screening ($\varepsilon_b=25$).} \label{q_disp}
\end{figure}

In Fig.~\ref{q_disp} we show the (log-scaled) $\text{EELS}(q,\omega)$ of the SSH model for different parameter sets. In Fig.~\ref{q_disp}(a) ($t\neq t'$ and $\varepsilon_b=1$) we observe two plasmon branches corresponding to the in- (LPC) and out-of-phase (HPC) charge oscillations within the unit cells. The two branches are separated by a plasmonic energy gap, which closes when the single-particle gap closes (i.e. $t=t'$), as depicted in Fig.~\ref{q_disp}(b). At the same time, this plasmonic gap is also strongly affects by the Coulomb interaction, as illustrated in Fig.~\ref{q_disp}(c) where we use the same $t$ and $t'$ as in the Fig.~\ref{q_disp}(a) but increased dielectric screening ($\varepsilon_b=25$). In such case, the plasmonic gap greatly shrinks and the two branches nearly touch. 

\section{\label{Append_Stability} Plasmonic Excitation Energy \& Localization Stability}

To understand our observations from section~\ref{sec4-2Stability}, we go back the decomposition of the full charge susceptibility and the definition of $\bm{\chi}_0^\text{topo}$. p-IPR$^\text{topo}$ and EELS$^\text{topo}(\omega)$ are derived from the perturbed $\bm{\tilde{\varepsilon}}^\text{topo} = \bm{\mathds{I}} - \bm{V_c} \bm{\tilde{\chi}}_0^\text{topo}$. The only quantity within this expression which is modulated by the applied perturbation is $\bm{\tilde{S}}(\omega)$, which can be approximated by $\bm{\tilde{S}}(\omega) \approx c(a) \bm{S}(\omega)$, where $c(a)$ is a scaling factor depending on the perturbation strength $a$. Thus $\bm{\tilde{\varepsilon}}^\text{topo}$ is affected just by a simple scaling factor which does not change its eigenvector, but its eigenvalue. Correspondingly, the excitation energy [EELS$^\text{topo}(\omega)$] is affected by the perturbation, but the real-space localization [p-IPR$^\text{topo}$] is not.

For the full quantities, i.e., p-IPR$^\text{full}$ and EELS$(\omega)$ derived from the perturbed $\bm{\tilde{\varepsilon}} = \bm{\mathds{I}} - \bm{V_c} \bm{\tilde{\chi}}_0$ this line of argumentation does not hold anymore. Here, $\bm{\tilde{\chi}}_0$ cannot be described as a scaled version of $\bm{\chi}_0$, so that both eigenvectors and energies are affected by the perturbation. Thus, $\bm{\chi}_0^\text{bulk}$ is responsible for the deloclaization of the topological plasmon with increasing perturbation, while $\bm{S}(\omega)$ is the main reason for the destabilization of its excitation energy.

\section{\label{Append_GlobalCoulomb}Global Variation of the Coulomb Interaction}

\begin{figure}[h]
 \includegraphics[width=8.6cm]{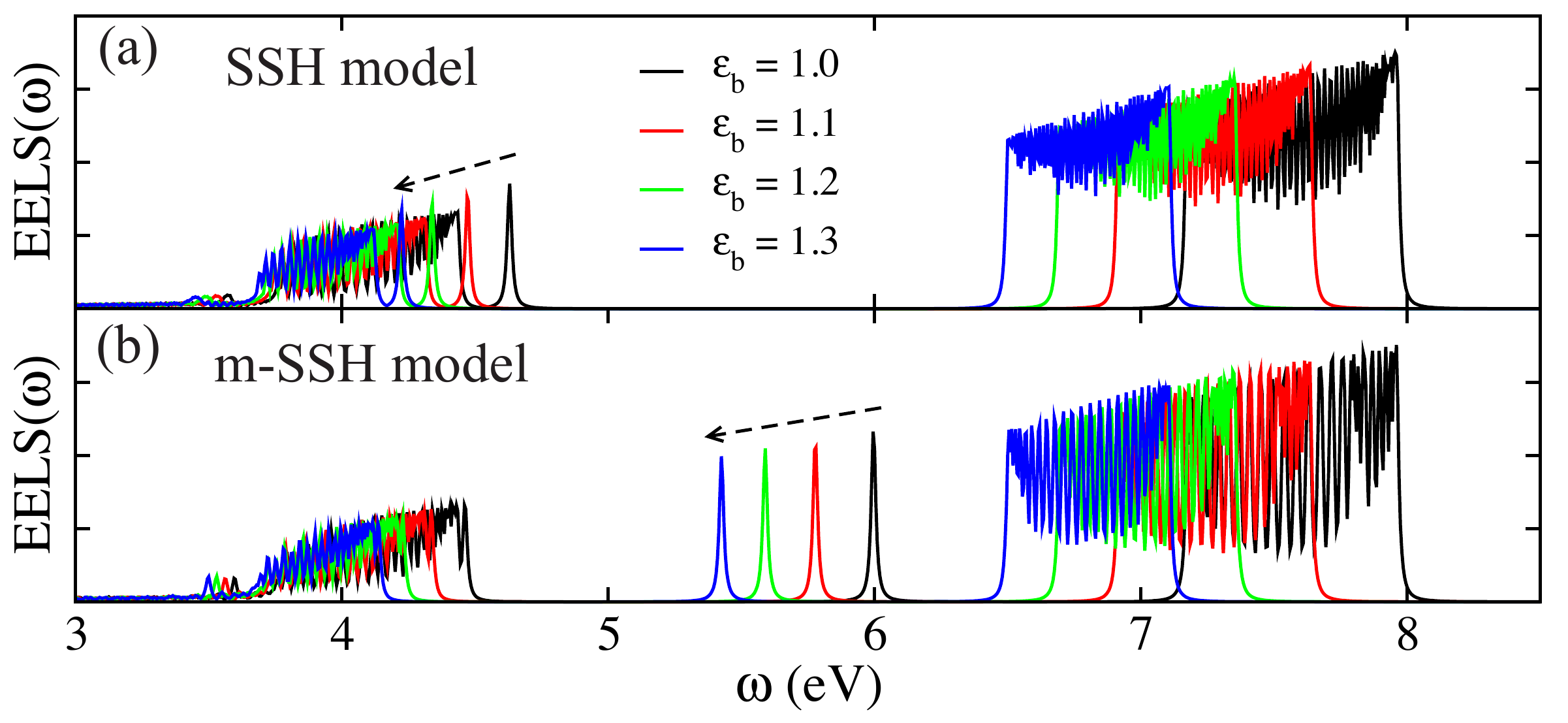}
 \caption{Effect of the screening environment on the plasmon spectrum. Shown are the EELS of (a) the open-ended 100-site topological SSH chain and (b) the open-ended 103-site m-SSH chain with strong interface, subject to different background dielectric constant $\varepsilon_b$.} \label{globalCoulomb}
\end{figure}

Globally varying the Coulomb interaction by changing the background dielectric constant $\varepsilon_b$ will shift the excitation energies of all plasmons including topological ones. We demonstrate this by showing the EELS$(\omega)$ of the SSH and m-SSH models in their topological non-trivial phase for varying $\varepsilon_b$ in Fig.~\ref{globalCoulomb}. The excitation spectra shift continuously to lower energies with increasing $\varepsilon_b$ and thus decreasing Coulomb interactions. In more detail, high-energy modes shift stronger than those with lower excitation energies.

%******** bibliography ********
\nocite{*}
\bibliographystyle{apsrev4-2}
\bibliography{bibs_all}% Produces the bibliography via BibTeX.

\end{document}